\documentclass[epj]{svjour}

\usepackage{graphicx}
\usepackage{graphics,epsfig,psfig}
\usepackage{bm}
\usepackage{amssymb}

\begin{document}


\title{ Structural transitions in  granular packs: statistical mechanics and statistical geometry investigations }


\author{ T. Aste and T. Di Matteo}


\institute{Department of Applied Mathematics, The Australian National University, 0200 Canberra, ACT, Australia. }


\abstract{
We investigate equal spheres packings generated from several experiments and from a large number of different  numerical simulations. 
The structural organization of these disordered packings is studied in terms of the network of common neighbours. 
This geometrical analysis reveals sharp changes in the network's clustering occurring at the packing fractions (fraction of volume occupied by the spheres respect to the total volume, $\rho$) corresponding to the so called Random Loose Packing limit (RLP, $\rho \sim 0.555$) and Random Close Packing limit (RCP, $\rho \sim 0.645$). 
At these packing fractions we also observe abrupt changes in the fluctuations of the portion of free volume around each sphere.
We analyze such fluctuations by means of a statistical mechanics approach and we show that these anomalies are associated to sharp variations in a generalized thermodynamical variable which is the  analogous for these a-thermal systems to the specific heat in thermal systems.
}

\titlerunning{Structural transitions in  granular packs}
\authorrunning{ T. Aste and T. Di Matteo}
\maketitle

\section{Introduction}

Since the earliest studies of granular materials it has been evident that one of the key quantities which affects the system's properties is the packing fraction (fraction of the total volume occupied by the grains).
It is well known since ancient times that different actions and different tunings of a given action can generate packings with different packing fractions.
Typically,  in experiments, such `actions' consist in tapping the system with vertical vibrations or  by shearing or by rotating the container or by pouring the grain in a container. 
In times when grain was sold by volume, the preparation protocol to achieve a dense packing was very important \cite{Bernal64}.
This is even reported  in the gospel as an example of \emph{good measure}:  ``Give, and it shall be given to you.  Good measure, pressed down, shaken together, running over, will be  put into your lap.''  (Luke 6:38).
Indeed, depending on the system handling, one can have variations in packing fractions up to 15\% within the two  limits $\rho \sim 0.555$ and $\rho \sim 0.645$ which are commonly refereed as Random Loose Packig (RLP) and Random Close Packing (RCP) limits.
For instance, by using the fluidized bed technique \cite{Schroder05,AsteEPL07} one can obtain packing fractions in the whole spectrum from 0.555 to 0.645 by varying the intensity of the flow pulses.

One of the scientists who first investigated the microscopic nature of granular packing was J.D.~Bernal that in a stream of papers concerning the ``structure of liquids'' reported some of the most important features of the structural organization of disordered sphere packings \cite{Bernal64,Bernal59,Bernal60}.
It was Bernal who pointed out  that disordered packing of equal spheres cannot overcome the RCP limit.
Fascinated by the simultaneous simplicity and complexity of these systems, he asked the following question:  ``Science is measurement, but what is a good measure?''.
Indeed, he was aware that depending on the kind of external actions  the system will result in different  packing fractions. 
However, he also observed that for a given external driving the system produces configurations with very similar and reproducible packing fractions which fluctuate in a very narrow range of  $0.5\%$. 

In this paper we show that the study of the fluctuations of such reproducible packing fractions can shed light on the origin of the RLP and RCP limits. 
Indeed, from a statistical mechanics perspective such fluctuations are a measure of the way in which the system is exploring the accessible phase-space under a given external driving.
The study of these fluctuations gives therefore insights about the accessible phase-space under given constraints. 
Granular materials are particle systems in which the sizes of the constituents are large enough such that they are not subject to thermal motion fluctuations. 
Therefore, a direct application of a thermodynamical theory is not straightforward.
However, in recent years several extensions of classical statistical mechanics approaches have been proposed for these systems   
\cite{Edwards89,Mehta89,Barrat01,Fierro02,Makse02,Ojha04,Richard05,Corwin05,Ciamarra06,Lechenault06}. 
In this paper we use a statistical mechanics approach to relate the packing fraction fluctuations with changes in the system's structural organisation and to understand the nature of the structural transitions occurring at RLP and RCP limits.
In his ``Bakerian lecture'', Bernal explained that in such systems there are two fundamental questions to be addressed: 
1) ``What is the structure?" and 2) ``Why has it got this structure?''.
And he resolved that the answer must be searched by two means: 
1) \emph{Statistical Geometry} and 
2) \emph{Statistical Mechanics}. 
By following his footsteps, in this paper we use a geometrical analysis and a statistical mechanics approach to understand what is happening at the two RLP and RCP limits.

\section{Materials and Methods}
\label{s.0}
\subsection{Experiments}
The experimental results reported in this paper concern experiments from the AAS database of disorder packings \cite{Database}.
Specifically, we investigate the six samples `A-F' described in details in \cite{AstePRL06,AstePRE05,AsteKioloa}.
They are dry packings of acrylic mono-sized spherical beads prepared with different methods in a range of packing fractions between 0.58 and 0.64.
The samples B, D-F contain approximately 35000 beads of diameter 1.59~$mm$.
Whereas, the samples A, C contain approximately 120000 beads of diameter 1.00~$mm$.
Polydispersities are within 0.5\% and they are placed inside a cylindrical container  with an inner diameter of $ 55\; mm$ and filled  to a height of $\sim 75\; mm$   \cite{AstePRL06,AstePRE05,AsteKioloa}.
We also report data from 12 experiments concerning glass beads in water prepared at packing fractions between 0.56 and 0.60 by means of a fluidized bed technique \cite{Schroder05,AsteEPL07}.
Each sample consists of about 145000 beads of diameter 250~$\pm$~13~$\mu m$ placed in a cylindrical glass  container with an inner diameter of 12.7~$mm$.

\subsection{Numerical simulations} 
We  generate a set of packings by means of  an event-driven molecular dynamic simulation of hard spheres  which uses a modified Lubachevsky-Stillinger algorithm \cite{Lubachevsky90,Donev05b,Skoge06}.
The algorithm starts from random points in space and makes them grow uniformly into non-overallping spheres with the sphere positions evolving in time  according to Newtonian dynamics.
The simulation is ended when the sphere sizes cannot be increased any longer and a `jammed' state with diverging collision rate is reached.
Large expansion rates produce jammed configurations with low packing fractions whereas slower growth rates lead to larger packing fractions. 
With this technique, the least dense attainable jammed configurations have packing fractions $\rho \sim 0.56$ which correspond to the RLP limit.
On the other hand, for very slow rates, crystalline nuclei with large packing fractions (up to the limit $\sim$0.74) can be formed.  
In our simulations, by varying the growth rate between 500 and 0.00001, we generate jammed configurations with packing fractions between 0.56 and 0.65.
We  also generate non-jammed configurations in the range of packing fractions between 0.1 and 0.55 by keeping the growth rate at 0.001 and arresting the simulation once the desired packing fraction is reached.
These non-jammed systems are packing models of (mostly) non-touching spheres placed in space without overlaps.
Clearly, they are not mechanically stable. 
They cannot be observed in experiments under gravity but they might be relevant in studies concerning micro-gravity experiments or in colloid suspensions with matching liquid density.
Some simulations with a modified Jodrey-Tory algorithm \cite{Jodrey85} have been also performed.
The algorithm starts from a set of overlapping spheres with repulsive interactions. 
It reduces overlaps, until all are removed, by moving spheres and gradually shrinking their radii \cite{Anikeenko08}. 
All numerical simulations use a cubic box with periodic boundary conditions.
All the analysis of static packing properties have been performed on numerical samples containing 10000 spheres.
Global packing fluctuations have been studied on systems of different sizes (from 200 to 10000 spheres) and repeating simulations several times (200 at least per each average packing fraction).

\section{Geometrical study of a structural transition}
\label{s.1}
It has been pointed out in \cite{Aste05rev} that the `common-neighbor analysis of structure', first introduced by Clarke and J\'ons\-son in \cite{Clarke93}, is a very powerful method to detect structural organization. 
Such a construction consists in considering couples of neighbouring  spheres  which stay with centers within a given threshold radial distance and retrieve all the neighbors that the two spheres have in common. 
In this paper we choose a threshold distance of $1.255$ sphere diameter.
Such a distance coincides with the one used by Hales to individuate `near' spheres  in his recent proof of the Kepler's conjecture \cite{ppp} and it was also used recently by \cite{Anikeenko08,Anikeenko07} in the geometrical study of the RCP limit.
This is a rather convenient distance: not too small in order to be little sensitive to local rearrangements, and not too large, in order to avoid the averaging out of local properties.
Let us remark that the choice of such threshold is not critical.
The properties reported in this paper (Fig.\ref{f.phase}) are consistently observed in a range of thresholds from 1.05 to 1.4.

\begin{figure} 
\centering
{\includegraphics[width=0.95\columnwidth]{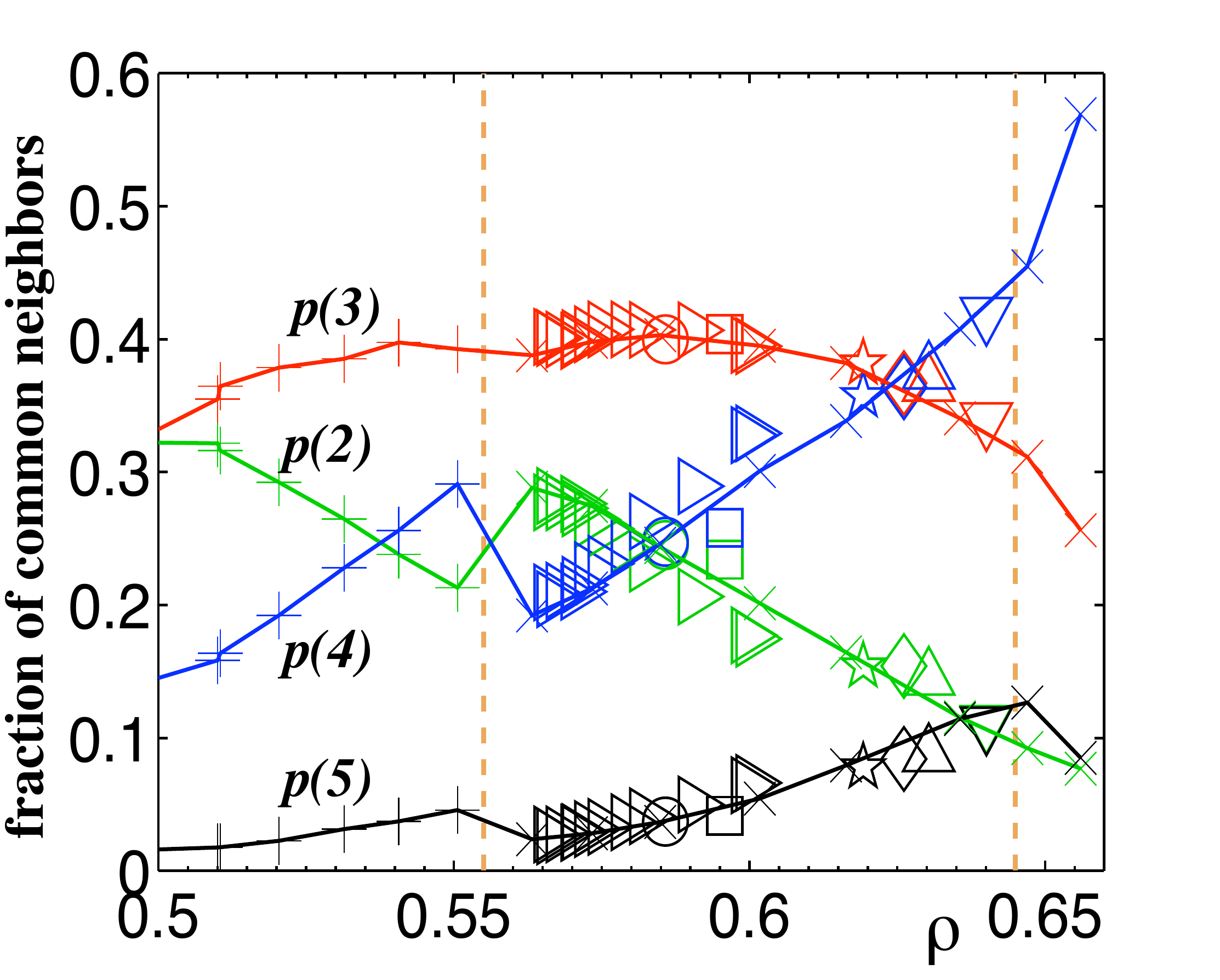}}
\caption{\footnotesize 
Fraction of common neighbors calculated from: 
1)~packing models of un-jammed packings of spheres (`$+$'); 
2)~jammed packings simulated by using a modified Lubachevsky-Stillinger algorithm \cite{Lubachevsky90,Donev05b,Skoge06} with different growth rates (`$\times$' ); 
3)~six experiments A-F \cite{AstePRL06,AstePRE05,AsteKioloa} with dry acrylic beads (`$\circ$, $\Box$, $\star$, $\diamond$, $\bigtriangleup$, $\bigtriangledown$'); 
4)~twelve experiments with glass spheres in water prepared by means of fluidized beds technique  \cite{AsteEPL07} (`$\rhd$'). }
\label{f.comm}
\end{figure}

\begin{figure} 
\centering
{\includegraphics[width=0.95\columnwidth]{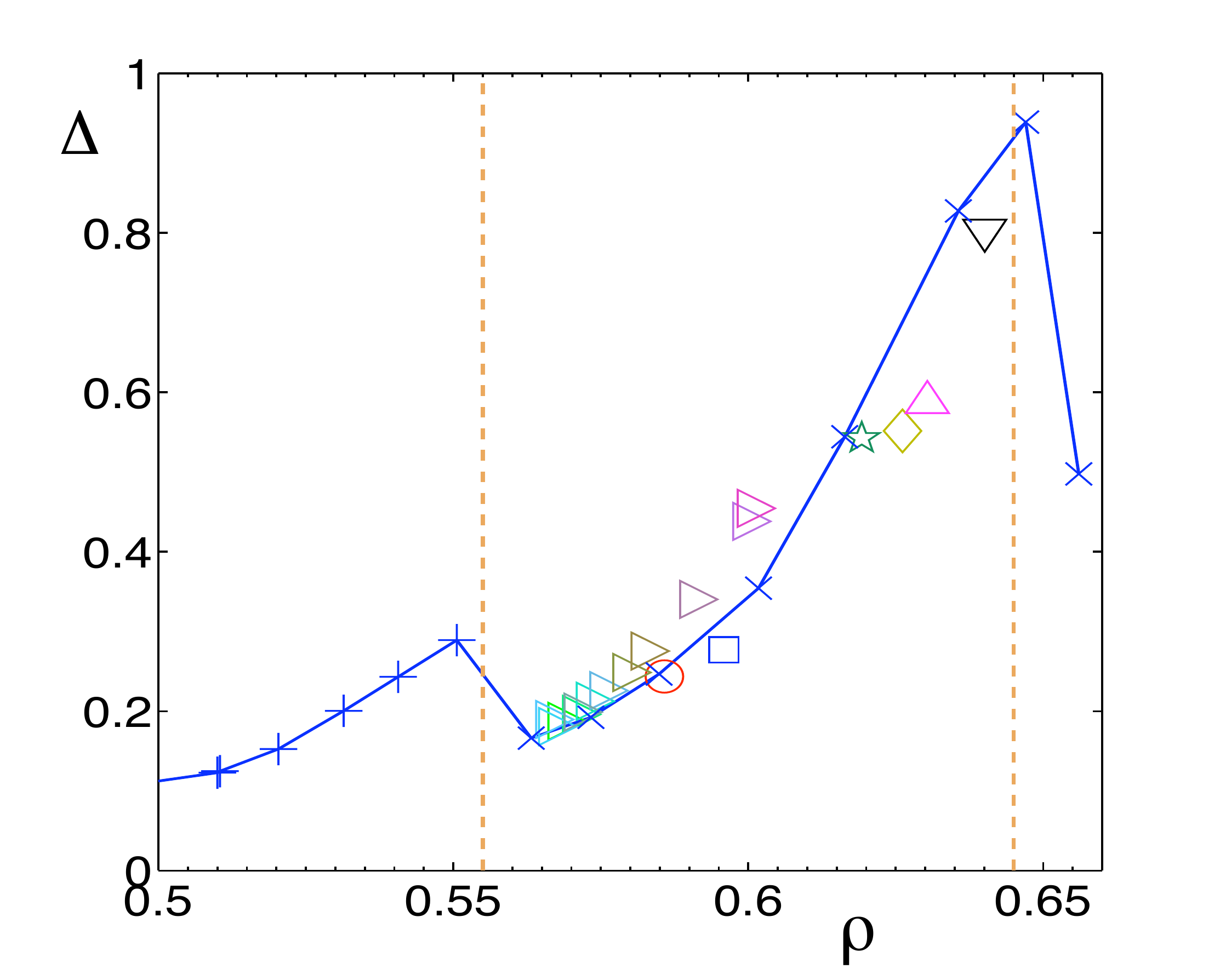}}
\caption{\footnotesize 
The ratio $\Delta= (p(3)p(5))/(p(2)p(4))$ vs. $\rho$ reveals very sharp changes at the packing fractions corresponding to the Random Loose Packing and Random Close Packing limits (the two vertical lines).
Symbols are as in Fig.\ref{f.comm}.
}
\label{f.phase}
\end{figure}

We calculate the fractions $p(q)$ of couples with $q$ common neighbors respect to the total number of couples. 
Figure \ref{f.comm} shows the results of such analysis performed on 15 numerical  simulations (modified Lubachevsky-Stillinger algorithm), 6 experiments with acrylic beads in air (A-F) and 12 experiments with glass beads in water (fluidized bed technique).
Only fractions with $q=2,3,4,5$ (which have the largest statistical weight) are shown.
The figure reveals a very good agreement between  simulations and experiments indicating that the structural properties are little sensitive to the preparation method and to the physical characteristics of the grains.
Conversely the figure reveals a clear universal dependence of the structural properties on the packing fraction $\rho$ with opposite trends for $p(2)$, $p(3)$ and $p( 4)$, $p(5)$.

A very clear signature that something is occurring to the structure around the Random Loose Packing packing fraction is revealed by the sharp changes in the behaviors of $p(2)$ and $p(4)$  occurring between  $\rho = 0.55$ and $\rho = 0.56$. 
Some evidences of the onset of a different regime can also be observed from the behavior of $p(3)$ and $p(5)$ around the  Random Close Packing limit ($\rho \sim 0.645$).
Such a change must be due to the fact that in the crystalline phase (which, in the simulated samples, begins to nucleate  above the RCP  limit) there are no configurations with $3$ or $5$ common neighbors.
Indeed, any closed packed  phase, made by stacking  hexagonal layers of spheres (Barlow packings \cite{ppp}), can admit only 2 or 4 common neighbors.

In order to better visualize any structural transition, in Fig.\ref{f.phase} we plot the ratio:
\begin{equation}
\Delta = \frac{p(3)p(5)}{p(2)p(4)} \;\;,
\label{delta}
\end{equation}
which is a `signature' of disordered arrangements weighting configurations which are impossible in crystalline packings (3 and 5 common neighbors) against configurations that instead are common in crystals  (2 and 4 common neighbors).
The plot of $\Delta$ reveals two sharp transitions occurring respectively at the RLP and at the RCP limits.

We have therefore acquired a first evidence that there  are neat structural changes occurring at the RLP and RCP limits.
Now let us establish if these geometrical changes are associated with other changes in the statistical mechanics properties of these systems.

\section{A statistical mechanics study }
\label{s.2}

As discussed in the introduction,  different preparation procedures and different experiments can result in granular packings with different total occupied volumes $V$ (or equivalently different packing fractions $\rho = \pi N d^3/(6 V)$ with $N$ the number of spheres and $d$ their diameters) . 
Here we are interested in the properties of the set of all possible total volumes (packing fractions) which can be reached by means of a chosen system's driving.

Any statistical mechanics theory will typically yield to an expression for the  probability distribution of the volume fluctuations at equilibrium of the following form \cite{AsteEPL07,AsteKGammaPRE08}: 
\begin{equation}
p_\infty(V) = \frac {  \Omega(V) e^{-   V/\chi   }   }{  \sum_{V'} \Omega(V') e^{- V '/\chi   } }\;\;\;;
\label{e.ME}
\end{equation}
where $\Omega(V)$ is the number of microscopic states which are classifiable under the same (coarse grained) state with vo\-lu\-me $V$.
The quantity $ \chi^{-1}$ is a temperature-like intensive variable which was named `compactivity' by Edwards \cite{Edwards89}.
It is determined by the  constraint on the average volume:
\begin{equation} \label{e.VmeL}
\bar V = \left< V \right> =  \sum_{V} V p_\infty(V)  \;\;.
\end{equation}
A derivation of Eq.\ref{e.ME} from a minimal set of statistical arguments is provided in \cite{AsteEPL07}; whereas a complete deductive statistical mechanics derivation is given in \cite{AsteKGammaPRE08}.

The challenge is to compute the number of equilibrium configurations $ \Omega(V) $ associated  with states which occupy a total volume $V$ under a given system preparation. 
To this end we can image that the whole system is made of a number $k$ of `elementary cells'  $\{{\mathbf c}_1,...,{\mathbf c}_k\}$ \cite{AsteKGammaPRE08,AsteEPL07}.
Let us stress that the number of such elementary cells does not  coincide in general with the number of grains in the system.
Given such cellular partition, $ \Omega(V) $  can be computed exactly, under the two following assumptions:
 (1) these cells can have arbitrary volumes above a minimum value  $v_{min}$, under the sole condition that the whole system must occupy a total volume $V$; 
 (2) all the cell-properties ${\mathbf c}_i$ are either completely determined by their volumes $v_i$ or they are independent from  $v_i$.
In this case, we have
\begin{eqnarray}
&&\Omega(V) =  \left(\frac{1}{\Lambda^3}\right)^k \int_{v_{min}}^V dv_1 \int_{v_{min}}^V dv_2 .... \int_{v_{min}}^V dv_k 
\nonumber \\
&&\delta(v_1+v_2+...+v_k - V)=   \frac{ (V- k v_{min} )^{k-1} }{\Lambda^{3 k}(k-1)!} \;\;\;,
\label{e.S1Dbb}
\end{eqnarray}
with $\Lambda$ a constant analogous to the Debye length. 
Substituting into Eq.\ref{e.ME}, and by using Eq.\ref{e.VmeL} we obtain 
\begin{equation}
\chi =\frac{(\bar V - k v_{min})}{k} \;\;\;;
\end{equation} 
and  
\begin{eqnarray}
&&p_\infty(V) = f(V,k) = \nonumber \\
&& \frac{k^{k}}{\Gamma(k) } \frac{(V - V_{min})^{(k-1)} }{(\bar V - V_{min})^{k}} \exp \left( {-k \frac{V-V_{min}}{\bar V - V_{min} } } \right)\;\;\;,
\label{e.pV1D}
\end{eqnarray}
with $V_{min} = k v_{min}$.
The function $f(V,k)$ is the probability density function to find a packing of $k$ elementary cells occupying a volume $V$ when the system is subject to an external driving that produces an average occupied volume $\bar V$.
Note that Eq.\ref{e.S1Dbb} is valid for any $k$. 
Indeed, the observable system  can be any arbitrary  sub-set of a larger system.
Moreover, the experiment can be performed either on several different independent systems or --~equivalently~-- on several non-iteracting sub-sets of a large system.
Eq.\ref{e.pV1D} is a Gamma distribution in the variable $V- V_{min}$; it is characterized by a `shape'  parameter  $k$ and a `scale'  parameter $\chi$ \cite{Gamma}. 
In Ref.\cite{AsteKGammaPRE08} we named such distribution k-gamma distribution. 
For this distribution the average volume $\left< V \right>$  coincides with  $ \bar V$ and the variance is
\begin{equation}
\sigma^2_v = \frac{(\bar V-V_{min})^2}{k} \;\;\;.
\label{e.variance}
\end{equation}
This last relation is very useful because it provides a practical means to evaluate $k$ from a set of volume measurements: $k =  (\bar V  -V_{min})^2/\sigma^2_v$.

\section{Granular temperature, fluctuation relation and Specific heat}
\label{s.3}

Equation \ref{e.pV1D} predicts that the statistical distribution of the volume fluctuations depends only on the parameter $k$ which counts the number of elementary cells in the  system.
Let us better understand the physical and statistical mechanics meaning of such quantity.
Following Edward's ideas \cite{Edwards89,Mehta89}, in granular systems a `granular temperature' (compactivity $\chi$) can be inferred from an analogy with the thermodynamical relation $\beta =1/(k_BT) = \partial (Entropy)/\partial (Energy)$ \cite{Edwards89,Mehta89}, by susbstituting the volume to the role played by the energy in thermodynamical systems.
In the present approach we can write the `statistical entropy'  (or Gibbs entropy) for an ergodic set $Z$ characterized by and average volume $\bar V$, as \cite{AsteKGammaPRE08}:
\begin{equation}
S(Z)= - \sum_{V \in Z} p_\infty(V) \log p_\infty(V) +  \sum_{V \in Z} p_\infty(V) S(V) \;\;\;,
\label{e.SZ}
\end{equation}
which, in the notation used in this paper, becomes
\begin{equation}
S(Z) = k\left[ 1+  \ln\left(  \frac{\bar V - V_{min}}{k \Lambda^d }\right)\right]  \;\;\;,
\label{Entropy}
\end{equation}
leading to
\begin{equation}
 \beta_{gr} = \frac{ \partial S(Z)}{\partial \bar V} = \chi^{-1} = \frac{k}{ \bar V  - V_{min}}  \;\;\;.
\label{EdwComp}
 \end{equation}
The Edwards' compactivity $\chi = \beta_{gr}^{-1}$  \cite{Edwards89,Mehta89} is therefore the average free-volume per elementary cell $\chi = (\bar V  - V_{min})/k$.
This means that, in the present approach, the `granular temperature' is a measure of the kind and the degree of space-partition into elementary cells.
The volume fluctuations within the ergodic set can be directly calculated from Eq.\ref{e.pV1D}  and one can verify that the correct relation between compactivity and volume fluctuations is attained:
 \begin{equation}
\chi^2 \frac{\partial \left< V \right>}{\partial \chi}  =  \left< (V - \left< V \right>)^2\right>   = \sigma_v^2 \;\;\;.
\label{e.fluct}
\end{equation}

From this equation, substituting Eq.  \ref{e.variance},  we obtain the following relation for the parameter $k$:
 \begin{equation}
k =  \frac{\partial \left< V \right>}{\partial \chi}   \;\;\;.
\label{e.k}
\end{equation}
The parameter $k$ measures therefore the amount of volume that must be added to the system in order to increase of one `granular degree' the compactivity.  
The analogous quantity for molecular gasses is: $\partial E/\partial T$, which is the \emph{specific heat}.
In analogy with ordinary thermodynamics such `specific heat' is expected to be sensitive to changes in the system's internal properties.

\begin{figure} 
\centering
{\includegraphics[width=.45\textwidth]{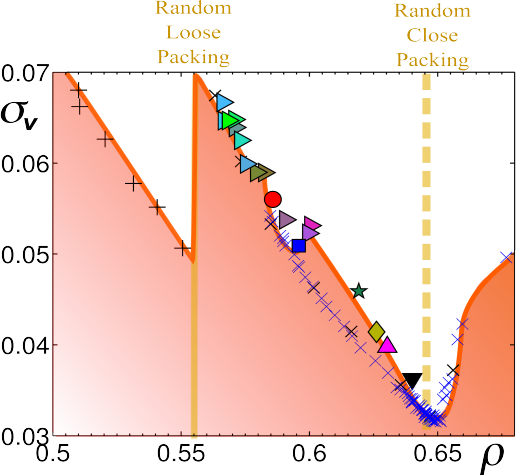}}
\caption{\footnotesize 
Behavior of the standard deviation of the Vorono\"{\i} volume fluctuations ($\sigma_v$) vs. packing fraction ($\rho$).
Very sharp changes are observed at $\rho \sim 0.555$ and at $\rho \sim 0.645$.
Symbols are as in Fig.\ref{f.comm}.
}
\label{f.sigrho}
\end{figure}

\begin{figure} 
\centering
{\includegraphics[width=.45\textwidth]{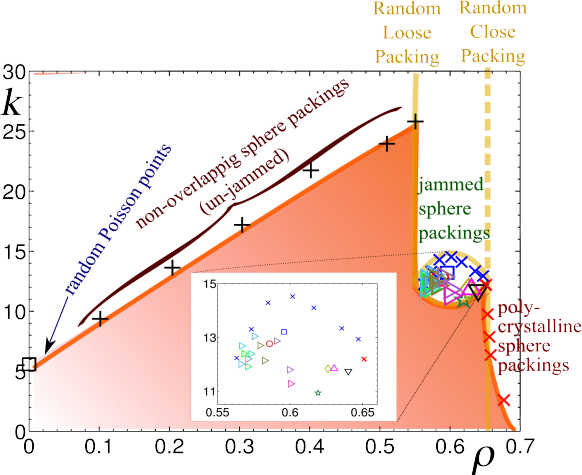}}
\caption{\footnotesize 
Behavior of $k$ calculated from the fluctuations of the Vorono\"{\i}  volumes by using $k =  (\bar V  -V_{min})^2/\sigma^2_v$ (Eq.\ref{e.variance}) vs. packing fraction $\rho$.
Symbols are as in Fig.\ref{f.comm}.
}
\label{f.krho}
\end{figure}

\begin{figure} 
\centering
{\includegraphics[width=.45\textwidth]{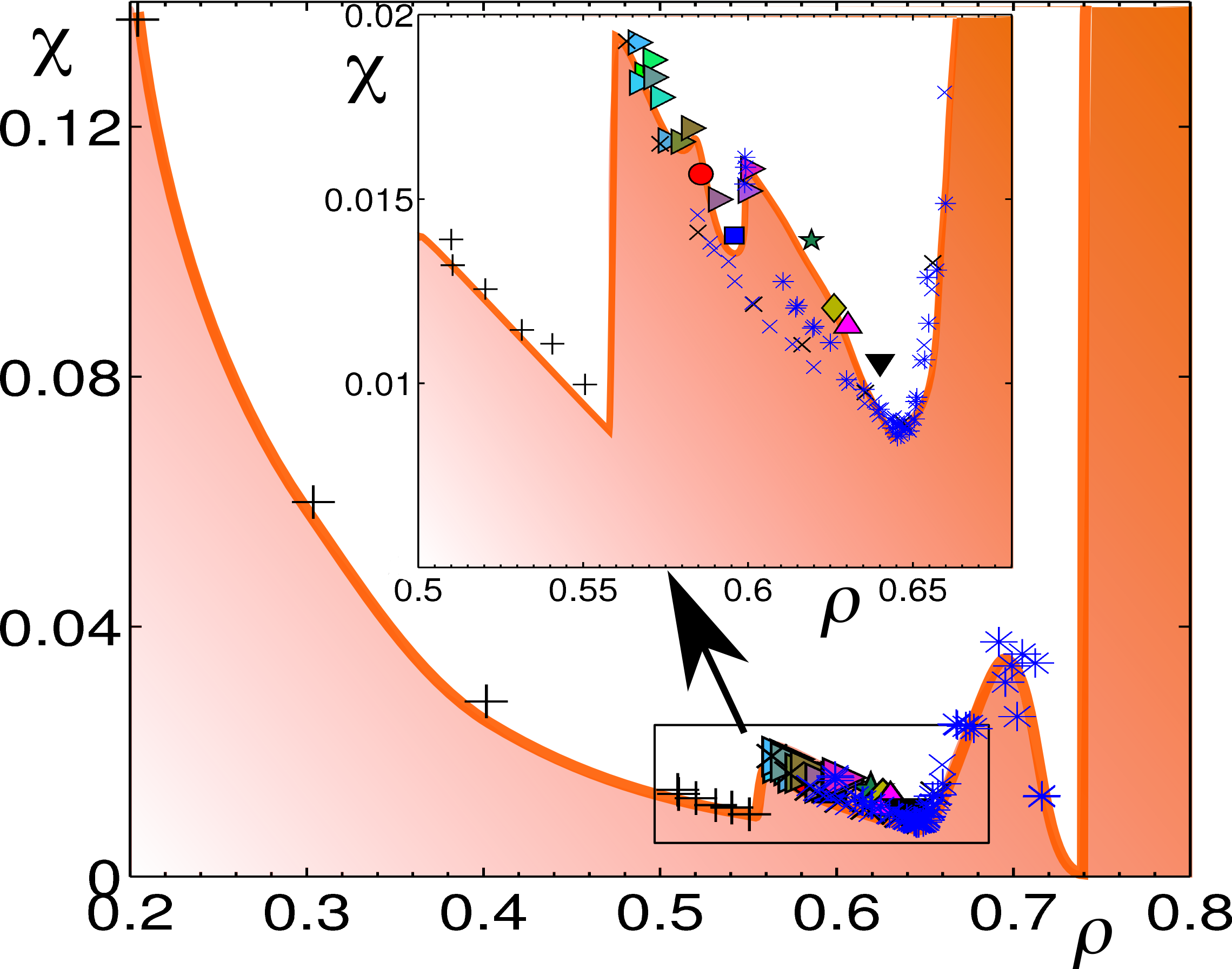}}
\caption{\footnotesize 
'Granular temperature' (compactivity  $\chi$)  vs. packing fraction $\rho$.
The symbols $*$ refer to 25 numerical simulations of packings with 10000 spheres generated by using the Jodrey-Tory algorithm \cite{Jodrey85}.
}
\label{f.chirho}
\end{figure}

\begin{figure} 
\centering
{\includegraphics[width=.45\textwidth]{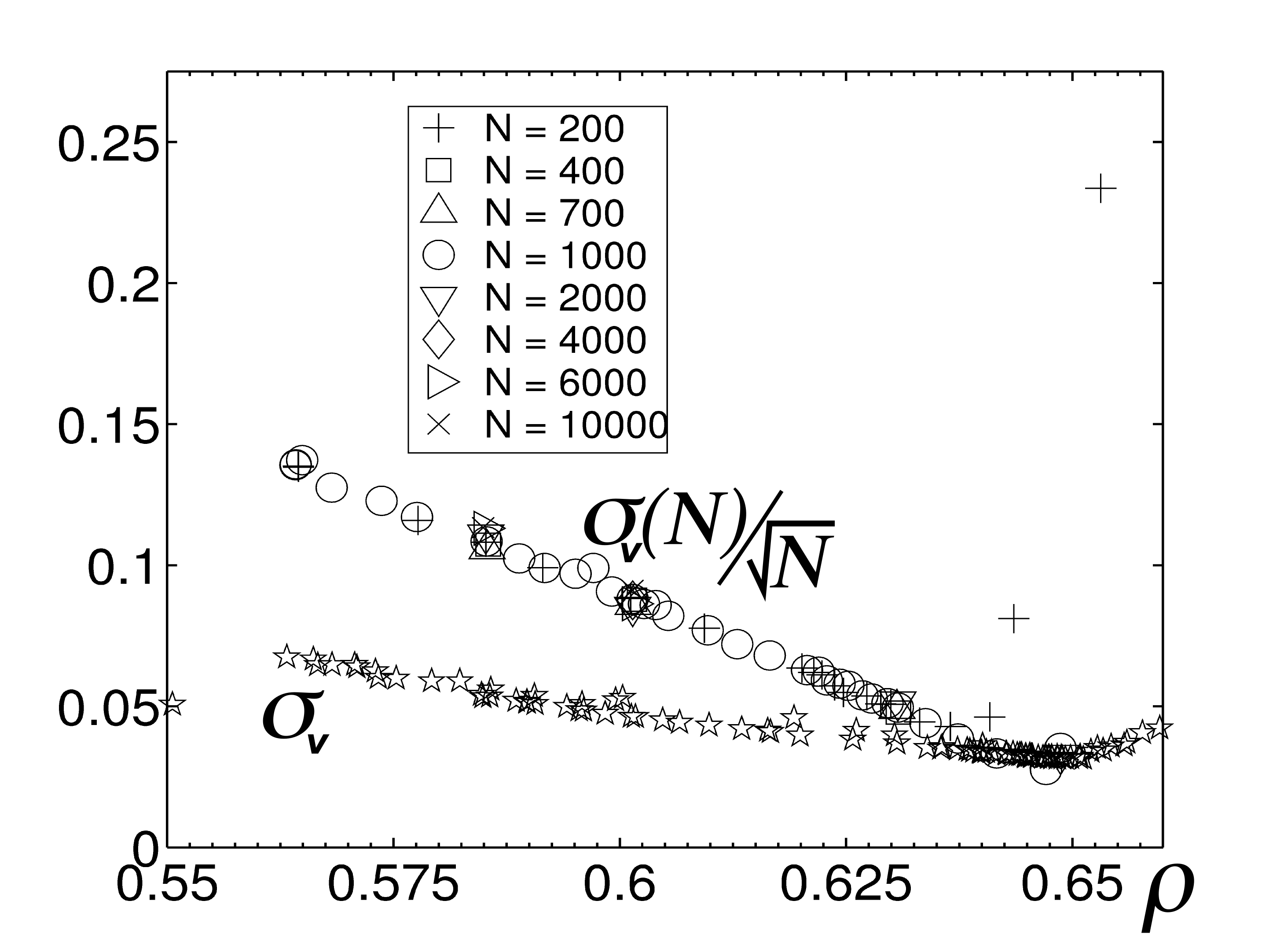}}
\caption{\footnotesize 
Rescaled standard deviations of the \emph{global} fluctuations of the whole sample volume $\sigma_v(N)/\sqrt{N}$ for various system sizes ($N$) calculated from several thousands numerical simulations  by using the Lubachevsky-Stillinger algorithm \cite{Lubachevsky90,Donev05b,Skoge06}.
All data rescale on the same values, deviations are observed only for the smallest sample ($N=200$, $+$) near the RCP limit where the small size of the sample increases the probability of crystallization.
The symbols `$\star$' refer to standard deviation of the \emph{local} Vorono\"{\i}  volume fluctuations $\sigma_v$ (same data as in Fig.\ref{f.sigrho}).
}
\label{f.sigNrho}
\end{figure}

\section{Changes in volume fluctuations around RLP and RCP limits}
\label{s.4}

We first investigate the volume fluctuations at the level of a single grain. 
For this purpose we use the Vorono\"{\i} partition where we calculate the  portion of space closest to  a grain center  respect to any other centre in the packing.
In Fig. \ref{f.sigrho} we report the standard deviation $\sigma_v$ of the distribution of volumes of the Vorono\"{\i} regions inside the various experiments and numerical samples.
One can observe that the fluctuations change abruptly in correspondence of the  two RLP and RCP limits.
We also observe that, within an overall decreasing trend, there are small but sizable changes at intermediate packing fractions such as 0.58 ad 0.6.
Such behavior is reflected in the value of the parameter $k =  (\bar V  -V_{min})^2/\sigma^2_v$ (Eq.\ref{e.variance}) as reported in Fig.\ref{f.krho}.
Note that the minimum volume of a Vorono\"{\i} region in packings of equal sphere with unit diameters is a fixed value corresponding to the volume of a dodecahedral region: $V_{min} = 0.694...$ \cite{AsteEPL07}.
The value at zero packing fraction ($k = 5.586$) was calculated analytically for random Poisson points in three dimensions  \cite{Gilbert62,Pineda04}.
Figure \ref{f.krho} shows that at low packing fractions, for non jammed configurations,  the value of $k$ increases almost linearly with $\rho$.
Then, it drastically decreases to values between 11 and 15 when the system gets into jammed configurations.
The inset in the figure shows that there are differences in the values of $k$ for different systems and within the same system at different packing fractions.
One can also note a rather sharp change in the experimental data occurring around $\rho \sim 0.6$ which might indicate some kind of transition at this packing fraction.
Above the packing fraction $\sim 0.645$ (RCP), the packings contain partially crystallized regions and the change in the kind of structural organization is reveled by a sharp drop in the value of $k$ that eventually will go to zero at the crystalline limit ($\rho = 0.740...$).
These data are consistent with Refs. \cite{AsteEPL07,AsteKGammaPRE08} where we have shown that the volume distribution of the Vorono\"{\i} regions follows remarkably well the theoretical prediction $f(V,k)$  (Eq. \ref{e.pV1D}) with $k$ in the range between $9 \le k \le 25$.
This implies that, depending on the kind of system in exam and on the packing fraction, there are between 9 to 25 elementary cells which are in average contributing to the volume of each Vorono\"{\i} region.
The impressive fact is that these systems are very different (ideal Newtonian spheres, acrylic beads in air and also glass beads in water \cite{AsteEPL07}) and they are prepared in very different ways (pouring, tapping, fluid flows, shearing, hard-spheres molecular dynamics). 
The fact that all these distributions follow the same law $f(V,k)$ (k-gamma distribution, Eq.\ref{e.pV1D}) suggests that there  are universal properties that determine the  packing configurations and their fluctuation laws.
On the other hand, the fact that different systems or different preparation methods yield to distributions with different values of $k$ indicates that this quantity is an important parameter to control and characterize the system's properties.
Equation \ref{EdwComp} reveals a direct relation between the parameter $k$ and the Edwards compactivity $\chi$.
The variation of the compactivity with the packing fraction is reported in Fig.\ref{f.chirho}.
One can observe that $\chi$ has a similar behavior to $\sigma_v$ (Fig.\ref{f.sigrho}) revealing sharp peaks within an overall decreasing trend.
Again we observe large changes occurring at the RLP and RCP limits indicating that strong changes in the system's properties are happening at these limits.
In the figure there are also reported data for packing models generated by using the Jodrey-Tory algorithm \cite{Jodrey85}.
With this algorithm we can reach larger packing fractions spanning a region above the RCP limit where the system becomes polycrystalline.

We also investigate the volume fluctuations at the level of the whole sample. 
In this case we can only study jammed configurations above the RLP limit.
In Ref.\cite{AsteKGammaPRE08} we demonstrated that the distribution of the total volume occupied by the packed spheres  follows accurately well the prediction of Eq.\ref{e.pV1D}.
The parameter $k$ calculated from the global fluctuations reveals a clear peak at the RCP limit (see Fig.6 in \cite{AsteKGammaPRE08}) wich is consistent with the abrupt changes at RCP observed in the local $k$ (Fig.\ref{f.krho}).
We observe that the global standard deviation of the volume distribution in a packing of $N$ spheres  ($\sigma_v(N)$) scales with the system size accordingly with the law: $\sigma_v(N) = \sigma_1 \sqrt{N}$.
This scaling law is clearly demonstrated in Fig.\ref{f.sigNrho} where $\sigma_v(N)/\sqrt{N}$ for various system sizes between $N= 200$ to $N= 10000$ all collapse onto a single trend.
Such a scaling confirms that the compactivity, calculated from the global fluctuations (Eqs.\ref{e.variance} and \ref{EdwComp}) $\chi = \sigma_v(N) / (\bar V - V_{min})$, is indeed an intensive parameter.
However, Fig.\ref{f.sigNrho} reveals that the scaling factor $\sigma_1$ does not coincide with the observed variance at the level of a single grain $\sigma_v$.
Such a discrepancy must be consequence of correlations between neighboring Vorono\"{\i} regions \cite{AsteScaling07}.
As consequence the compactivity measured at local level is different from the one measured from global fluctuations.  
In this respect, the `proper' compactivity is the one associated to the global volume fluctuations; the local measure is an `effective compactivity' \cite{AsteKGammaPRE08}.
Intriguingly, they coincide at the RCP limit.

\section{Conclusions}
\label{s.5}

In this paper we provide two independent evidences of transitions occurring in sphere packings at the Random Loose Packing and Random Close Packing limits.
\begin{itemize}
\item[(i)] The \emph{first} evidence is geometrical and it is acquired from the behavior of the  fraction of common neighbours.
In particular, we have observed that the fraction of couples with 3, 5 and 2 or 4 common neighbours (Eq.\ref{delta}) has sharp changes occurring both at the RLP and RCP limits. 
Such changes indicate that rearrangements towards a different packing organization  are occurring at these limiting packing fractions.
\item[(ii)] The \emph{second} evidence is acquired from a statistical mechanics study concerning the fluctuations in the local Vorono\"{\i}  volumes.
We have found that all the samples investigated follow the same kind of statistical distribution $f(V,k)$ (k-Gamma distribution, Eq. \ref{e.pV1D}) but they are characterized by different values of the quantity $k$. 
The value of $k$ sharply decreases when the packing fraction crosses the RLP or RCP limits.
This corresponds to sharp freezing of some degrees of freedom associated with changes in the packing' s organization.
We have discussed that the quantity $k$  is analogous to the specific heat in ordinary thermodynamics.
\end{itemize}
These evidences clearly demonstrate that transitions are occurring at the two RLP and RCP limits. 
However, it rests unclear whether such transitions can be described as `proper' phase transitions in a statistical mechanics framework.
This difficulty is intrinsically associated to the fact that the system is disordered, there are no symmetries to break  and we cannot introduce an `order parameter' to simply describe the structural changes. 
Nevertheless, we have clearly shown in this paper that abrupt changes in the structural properties are occurring and that such changes are associated with freezing of degrees of freedoms and consequent contractions of the available phase-space.
Specifically, at the  RLP limit the system undertake important changes passing from a `compressible gas' -like behavior to a `rigid solid' -like behavior. 
If we constraint our analysis to mechanically stable structure  (rigid or `jammed' structures only) then we see that both RLP and RCP limits are associated with exhaustion of realizable packings. 
Experimental preparation methods typically fail to find disordered packings with densities above $\rho_{RCP} \simeq 0.645$ even if there exists a large class of layered packings (Barlow packings \cite{ppp}) with packing fraction $\rho = 0.740...$.   
A study of the system entropy (Eq.\ref{Entropy}) reveals that the number of accessible configurations decreases approaching the RCP transition but it becomes of the order of one only at the estimated  `Kauzmann density'  $\rho_K \sim 0.66$ \cite{Anikeenko08} which is larger than the observed $\rho_{RCP}$.
Similarly, in the RLP case mechanically stable structures are not discovered in disordered arrangements below $\rho_{RLP} \simeq 0.555$.
On the other hand, below the RLP limit jammed packings can be obtained, but only in special crystals with self-avoiding ``tunnels''  \cite{TorquatoRLP07}.
This seems to indicate that outside the RLP-RCP limits there might be configurations but they are isolated regions or points in the phase-space and they cannot be simply reached from small improvements on the known solutions, they have therefore infinitesimal probabilities to be discovered. 

\subsection*{Acknowledgements}
Many thanks to T. Senden, M. Saadatfar, A. Sakellariou, A. Sheppard, A. Limaye for the tomographic data.
Many thanks also to M. Schr\"oter and  H. Swinney for the fluidized bead experiments and many discussions.
This work was partially supported by the ARC discovery project DP0450292.
We thank AAS (Genova, Italy) for providing infrastructure and hospitality.

\bibliographystyle{epj}

\begin{thebibliography}{34}

\bibitem{Bernal64}
J.D. Bernal, Proc. R.Soc. Lond. \textbf{A280}, 299 (1964)

\bibitem{Schroder05}
M.~Schr\"oter, D.I. Goldman, H.L. Swinney, Phys. Rev. E. \textbf{71}, 30301 (R)
  (2005)

\bibitem{AsteEPL07}
T.~Aste, T.~Di{~}Matteo, M.~Saadatfar, T.~Senden, M.~Schr\"oter, H.L. Swinney,
  Eur. Phys. Lett. \textbf{79}, 24003 1 (2007)

\bibitem{Bernal59}
J.D. Bernal, Nature \textbf{183}, 141 (1959)

\bibitem{Bernal60}
J.D. Bernal, J.~Mason, Nature \textbf{188}, 910 (1960)

\bibitem{Edwards89}
S.~Edwards, R.~Oakeshott, Physica A \textbf{157}, 1080 (1989)

\bibitem{Mehta89}
A.~Mehta, S.F. Edwards, Physica A \textbf{157}, 1091 (1989)

\bibitem{Barrat01}
A.~Barrat, J.~Kurchan, V.L. ad~M.~Sellitto, Phys. Rev. E \textbf{63}, 051301
  (2001)

\bibitem{Fierro02}
A.~Fierro, M.~Nicodemi, A.~Coniglio, Europhys. Lett. \textbf{59}, 642 (2002)

\bibitem{Makse02}
H.A. Makse, J.~Kurchan, Nature \textbf{415}, 614 (2002)

\bibitem{Ojha04}
R.P. Ojha, P.A. Lemieux, P.K. Dixon, A.J. Liu, D.J. Durian, Nature
  \textbf{427}, 521 (2004)

\bibitem{Richard05}
P.~Richard, M.~Nicodemi, R.~Delannay, P.~Ribiere, D.~Bideau, Nature Materials
  \textbf{4}, 121 (2005)

\bibitem{Corwin05}
E.I. Corwin, H.M. Jaeger, S.R. Nagel1, Nature \textbf{435}, 1075  (2005)

\bibitem{Ciamarra06}
M.P. Ciamarra, M.~Nicodemi, A.~Coniglio, preprint.  (2006)

\bibitem{Lechenault06}
F.~Lechenault, F.~da~Cruz, O.~Dauchot, E.~Bertin, J. Stat. Mech.
  \textbf{P07009}, 1742 (2006)

\bibitem{Database}
See, http://wwwrsphysse.anu.edu.au/granularmatter/  (2006)

\bibitem{AstePRL06}
T.~Aste, Phys. Rev. Lett. \textbf{96}, 018002 (2006)

\bibitem{AstePRE05}
T.~Aste, M.~Saadatfar, T.J. Senden, Phys. Rev. E. \textbf{71}, 061302 (2005)

\bibitem{AsteKioloa}
T.~Aste, M.~Saadatfar, A.~Sakellariou, T.~Senden, Physica A \textbf{339}, 16
  (2004)

\bibitem{Lubachevsky90}
B.D. Lubachevsky, F.H. Stillinger, J. Stat. Phys. \textbf{60}, 561 (1990)

\bibitem{Donev05b}
A.~Donev, S.~Torquato, F.H. Stillinger, J. Comput. Phys. \textbf{202}, 737
  (2005)

\bibitem{Skoge06}
M.~Skoge, A.D.F.H. Stillinger, S.~Torquato, Phys. Rev. E. \textbf{74}, 041127
  (2006)

\bibitem{Jodrey85}
W.S. Jodrey, E.M. Tory, Phys. Rev. A. \textbf{32}, 2347 (1985)

\bibitem{Anikeenko08}
A.V. Anikeenko, N.N. Medvedev, T.~Aste, Phys. Rev. E \textbf{77}, 031101 (2008)

\bibitem{Aste05rev}
T.~Aste, J. Phys.: Condens. Matter \textbf{17}, S2361 (2005)

\bibitem{Clarke93}
A.S. Clarke, H.~J\'onsson, Phys. Rev. E \textbf{47}, 3975 (1993)

\bibitem{ppp}
T.~Aste, D.~Weaire, \emph{The Pursuit of Perfect Packing} (Institute of
  Physics, Bristol, 2000)

\bibitem{Anikeenko07}
A.~Anikeenko, N.~Medvedev, Phys. Rev. Lett. \textbf{98}, 235504 (2007)

\bibitem{AsteKGammaPRE08}
T.~Aste, T.~Di{~}Matteo, Phys. Rev. E \textbf{77}, 021309 (2008)

\bibitem{Gamma}
R.V. Hogg, A.T. Craig, \emph{Introduction to Mathematical Statistics}
  (Macmillan, New York, 1978)

\bibitem{Gilbert62}
E.N. Gilbert, Ann. Math. Stat. \textbf{33}, 958 (1962)

\bibitem{Pineda04}
E.~Pineda, P.~Bruna, D.~Crespo, Phys. Rev. E. \textbf{70}, 066119 1 (2004)

\bibitem{AsteScaling07}
T.~Aste, T.~Di{~}Matteo, Eur. Phys. J. E \textbf{22}, 235  (2007)

\bibitem{TorquatoRLP07}
S.~Torquato, F.H. Stillinger, J. Appl. Phys. \textbf{102}, 093511 (2007)

\end{thebibliography}

\end{document}